\documentclass[12pt]{article}
\usepackage{amsfonts}
\usepackage{amsmath}
\usepackage{epsfig}

\begin{document}

\begin{flushright}
10/2006\\
revised version 11/2006\\
\end{flushright}
\vspace{20mm}
\begin{center}
\large {\bf Tetrons - a possible Solution to the Family Problem}\\
\mbox{ }\\
\normalsize
\vskip3cm
{\bf Bodo Lampe}               
\vskip0.3cm
Institut f\"ur theoretische Physik \\
Universit\"at Regensburg \\
email: Lampe.Bodo@web.de \\   
\vspace{1.5cm}
{\bf Abstract}\\
\end{center}
A model is presented, in which fermion and vector boson 
states are constructed from constituents ('tetrons'). 
The model encodes all observed structures 
and phenomena of elementary particle physics in 
group theoretic items of the permutation group $S_4$. 
Details of the model like 
symmetry breaking, distribution of charges and 
mass generation are worked out.  
As a sideproduct a deeper understanding of parity 
violation is obtained. 


\newpage

\section{Introduction} 

According to present ideas the observed elementary particles (leptons, 
quarks and vector bosons) are pointlike. Their mathematical 
description \cite{gaugetheory} as Dirac or Yang-Mills fields follows 
this philosophy. In the present paper I propose a model, in which 
they acquire an extension and are composite of more fundamental fields 
called tetrons. 


\section{The Model for Fermions}

My aim is to shed light on the fermion spectrum of elementary 
particle physics, i.e. the 24 spin-$\frac{1}{2}$-states observed in nature 
and habitually denoted by 
\begin{equation} 
 \nu_e \qquad e \qquad u_{1,2,3} \qquad d_{1,2,3}  
\end{equation}
\begin{equation} 
 \nu_\mu \qquad \mu \qquad c_{1,2,3} \qquad s_{1,2,3}     
\end{equation}
\begin{equation} 
 \nu_\tau \qquad \tau \qquad t_{1,2,3} \qquad b_{1,2,3}  
\label{eq1}
\end{equation}
where the index 1,2,3 stands for quark color. 
These states arrange themselves in 3 families of 
\begin{equation} 
 N \qquad E \qquad U_{1,2,3} \qquad D_{1,2,3}  
\label{eq2}
\end{equation}
each consisting of two quadruplets of the form $(L,Q_{1,2,3})$ where I 
abbreviate quark states by the letter Q, up-quarks by U, down-quarks by D, 
leptons by L, neutrinos by N and e, $\mu$ and $\tau$ by E.  
Including spin (and righthanded neutrinos) there are 
altogether 48 degrees of freedom with a mass spectrum which ranges 
between about $10^{-6}$ eV for the electron-neutrino \cite{neutrinos}
and 175 GeV for the top-quark \cite{topquark}. 

The underlying dynamics of this system is nowadays usually 
supposed to be a local gauge theory with gauge group 
$U(1)_{B-L}\times SU(3)_c \times SU(2)_L \times SU(2)_R$ \cite{su2su2},  
$SU(4)\times SU(2)_L \times SU(2)_R$ \cite{su4} or $SO(10)$ \cite{so10}
with 15, 21 or 45 gauge bosons respectively plus an as 
yet unknown mechanism which takes care of the family repetition.  

I want to accomodate this system as bound states of smaller, more 
fundamental objects. The neatest idea \cite{lampeb} is to consider 
states consisting of 4 identical particles ('tetrons') bound 
together by a new super strong, super short range 
interaction, whose charge and symmetry nature are not 
discussed at this point but will become transparent in 
the course of discussion. 
The tetrons transform under the permutation group $S_4$ \cite{s4} 
and will be seen to yield the observed spectrum of quarks and 
leptons. 

In a crude classical picture they may be assumed to sit on the 
corners of an equilateral tetrahedron with corresponding 
symmetry properties, i.e. invariance under the tetrahedron's 
symmetry group $T_d$ which is isomorphic to the group $S_4$ of 
permutations of the set of the 4 tetrons which for convenience 
we numerate as 1,2,3,4. The group consists of 
24 permutations which may be denoted as 
$\sigma=\overline{abcd}:(1,2,3,4)\rightarrow (a,b,c,d)$. 
In the limit of perfect $S_4$-symmetry the 24 tetrahedron 
states $\phi_{\sigma}=|abcd>$ generated by these 
permutations are of course all identical. 

In order to become different with different 
masses and charges there must be some sort of symmetry breaking, 
the simplest possibility being that the 4 originally identical 
tetrons together with their bound states differ from each other 
even when rotational symmetry 
is taken into account. In order to implement this there 
are several possibilities which will be described later. 
At the moment I simply assume that a symmetry breaking 
exists and want to show how the 24 different permutation states 
can be arranged in parallel to the observed family structure. 

The nonabelian group $S_4$ may be written as a semidirect product 
\begin{equation} 
S_4= Z_4 \diamond Z_3 \diamond Z_2  
\label{eq1sd}
\end{equation}
where $Z_n$ is the (abelian) symmetric group of n elements. 
The subgroup $Z_3 \diamond Z_2$ of $S_4$ can be identified with the 
permutation group $S_3$ of 3 elements, whereas $Z_4 \diamond Z_2$ is 
isomorphic to the dihedral group $D_4$.  
\footnote{All these groups have geometrical interpretations 
as symmetry groups of simple geometrical objects.  
For example, $D_4$ is the symmetry group of a square, 
$S_3$ of an equilateral triangle and $K$ that of a rectangle 
(everything in 2 dimensions). $S_4$ itself is isomorphic to 
the symmetry group of a unilateral tetrahedron in 3 dimensions 
and as such may be considered as a subgroup of SO(3). 
The latter fact, however, will not be relevant (because we are 
going to completely break $S_4$), until the point when 
parity violation of the weak interactions is discussed 
in the section on vector bosons.}

The $Z_3$-part of the decomposition (\ref{eq1sd}) allows to 
divide the 24 elements of $S_4$ into 3 groups of 8 
elements ('orbits'), which correspond 
to permutations which preserve a certain 
ordering (...1234123...+backward, ...2134...+backward and 
...4231...+backward) and will essentially make up the 3 fermion 
families. 

There are then two possibilities to associate the up-half $(N,U_{1,2,3})$ 
of a family to elements of $S_4$. One is based on the 
so-called $Kleinsche$ $Vierergruppe$ $K$ (also a subgroup of $S_4$). 
It is the smallest noncyclic group and isomorphic 
to $Z_2\times Z_2$. Considered as a subgroup of $S_4$ it 
consists of the 3 even permutations $\overline{2143}$, $\overline{3412}$, 
$\overline{4321}$, where 2 pairs of numbers are 
interchanged, plus the identity. 
The other possibility is to take 
the $Z_4$ in eq. (\ref{eq1sd}), i.e. to identify 
the up-half of the first family (where by $first$ I do 
not necessarily mean lightest, see later) 
with the 3 permutations $\overline{2341}$, 
$\overline{3412}$, $\overline{4123}$ plus the identity. 
The down-half is then associated to the 
'backward running' permutations $\overline{4321}$, 
$\overline{3214}$, $\overline{2143}$ and $\overline{1432}$.
In both cases one will have to look for a mechanism, which 
suppresses leptoquarks, i.e. unwanted lepton number violating 
processes like proton decay etc. This question will be 
followed in the section about vector bosons.

In table 1 there is a preliminary assignment of particle states 
to permutations of $S_4$. It should be noted, however, that this 
is only mnemonic, because it will later turn out that one 
has to use linear combinations of $S_4$-states. 

\begin{table}
\label{tab1}
\begin{center}
\begin{tabular}{|l|c|c|c|}
\hline
&...1234...&...1423...&...1243... \\
& family 1 & family 2 & family 3 \\
\hline
& $\tau$, $b_{1,2,3}$ & $\mu$, $s_{1,2,3}$ & $e$, $d_{1,2,3}$ \\
\hline
$F_0$ & $\overline{1234} (id)$ & $\overline{2314}$ & $\overline{3124}$ \\
$F_1$ & $\overline{2341} (z_1) $ & $\overline{3142}$ & $\overline{1243} (3\leftrightarrow 4)$ \\
$F_2$ & $\overline{3412} (k_2=z_2)$ & $\overline{1423}$ & $\overline{2431}$ \\
$F_3$ & $\overline{4123} (z_3)$ & $\overline{4231} (1\leftrightarrow 4)$ & $\overline{4312}$ \\
\hline
& $\nu_\tau$, $t_{1,2,3}$ & $\nu_\mu$, $c_{1,2,3}$  & $\nu_e$, $u_{1,2,3}$  \\
\hline
$B_0$ & $\overline{3214} (1\leftrightarrow 3)$ & $\overline{1324} (2\leftrightarrow 3)$ & $\overline{2134} (1\leftrightarrow 2)$ \\
$B_1$ & $\overline{4321} (k_3)$ & $\overline{4132}$ & $\overline{4213}$ \\
$B_2$ & $\overline{1432} (2\leftrightarrow 4)$ & $\overline{2413}$ & $\overline{3421}$ \\
$B_3$ & $\overline{2143} (k_1)$ & $\overline{3241}$ & $\overline{1342}$ \\
\hline
\end{tabular}
\bigskip
\caption{Preliminary list of elements of $S_4$ ordered 
in 3 families. F are the forward oriented elements (
corresponding to the $Z_4$ subgroup of $S_4$), and B the 
backward oriented ones. Together they give $D_4$. 
$z_i$ and $k_i$ denote the elements of $Z_4$ and $K$, 
respectively, and $(a\leftrightarrow b)$ a 
simple permutation where a and b are interchanged. 
An alternative assignment based on $K$ instead of $Z_4$ 
is possible where the rows $F_1 \leftrightarrow B_3$ 
and $F_3 \leftrightarrow B_1$ are interchanged. 
In any case permutations with a 4 at the last position form 
a $S_3$ subgroup of $S_4$ and may be thought of giving the set 
of lepton states. Note however, that the assignment of particle 
states is only mnemonic, because it will turn out that one has 
to use linear combinations of $S_4$-states.
Furthermore, there is still an arbitrariness as to whether 
the 'first' family is the lightest or the heaviest, i.e. 
whether the identity permutation $\overline{1234}$ corresponds 
to the top quark or to the electron and so on. This is simply due 
to the fact that I have not yet discussed the question of 
symmetry breaking, masses and charges, but will do so in section 4. 
}
\end{center}
\end{table}

Another way of representing the above classification is 
by saying that 24 elements of $S_4$ are divided in 
6 classes which transform under its subgroup $S_3$. In fact, 
$S_3$ consists of 3 even permutations denoted by $id=\overline{I,II,III}$, 
$g_1=\overline{III,I,II}$ and $g_2=\overline{II,III,I}$ 
and 3 odd permutations denoted by $u_0=\overline{III,II,I}$, 
$u_1=\overline{II,I,III}$ and $u_2=\overline{I,III,II}$. 
It can be further decomposed as $S_3 \approx Z_3\diamond S_2$ where 
$S_2=Z_2$ and $Z_3$=$\{id,g_1,g_2\}$ is the cyclic group of 3 elements and 
in geometrical terms corresponds to rotations by $\pm 2\pi/3$. 
This part of $S_3$ is assumed to account for the 3 observed 
fermion families, i.e. to be the family group while the $S_2$-part 
will correspond to weak isospin, i.e. flip neutrino and electron. 

The down-quadruplet $(E,D_{1,2,3})$ can then be represented by a 
single even element of $S_3$ and $(N,U_{1,2,3})$ by a single odd element, 
while the 4 states within a quadruplet correspond to the orbits of $Z_4$ 
in $S_4$. The situation is summarized in the following list:  
\begin{itemize}
\item 
1.family: path ...1234... corresponding to $id=\overline{I,II,III}$ 
(+backward $p_2=\overline{III,II,I}$)
\item 
2.family: path ...2134... corresponding to $p_1=\overline{II,I,III}$ 
(+backward $g_1=\overline{III,I,II}$)
\item 
3.family path ...4231... corresponding to $g_2=\overline{II,III,I}$ 
(+backward $p_3=\overline{I,III,II}$)
\end{itemize}
In geometrical terms the 3 families can be visualized as the 
3 closed paths which can be drawn in a tetrahedron. The interactions among 
the tetrons in the i-th family runs along the path i, i=I,II or III. 
A given state $\phi_{\sigma}=|abcd>$ has thus bindings along 
the (open) path $a\rightarrow b\rightarrow c\rightarrow d$. 

All observed Fermions have Spin $\frac{1}{2}$ and they have associated 
antifermions. These features are indispensable for any theory of 
elementary particles. How can they be accommodated in the present model? 
Concerning these questions I refer to a forthcoming paper \cite{lampxx}
where it is argued that 
\begin{itemize}
\item Assuming tetrons are given by complex scalar fields 
$\chi (x)$, anti-tetrons are described by $\chi^* (x)$. 
\item So far we have 24 tetrahedron states $\phi_{\overline{abcd}}$. 
Assuming a relativistic motion these states 
will appear in left and righthanded form $\phi^L_{\sigma}$ 
and $\phi^R_{\sigma}$. If the lefthanded live on a 
tetrahedron $\vec{T}$ the righthanded will be shown to 
live on the 'anti-tetrahedron' $-\vec{T}$ which 
is obtained from the original one by parity 
inversion $P:\vec{x} \rightarrow -\vec{x}$ and is in 
fact the tetrahedron which completes $\vec{T}$ to a cube. 
It is this cube, by the way, and its 
centre and axes, which will later be used to describe 
vector boson states (see section 5). 
Note that although most of the time I consider tetrahedrons 
at rest, the present framework will be suitable general to be 
applicable to relativistic tetrahedron states.  
It further turns out that states of even and odd 
permutations naturally live on opposite tetrahedrons 
$\vec{T}$ and $-\vec{T}$, so that it is in principle 
important to keep track of right and lefthanded 
components, as will be discussed in more detail in section 5. 
For simplicity of representation, however, I will not 
differentiate at the moment. 
\item The question of the spin of the constituent tetrons will be 
further discussed in \cite{lampxx}. 
\end{itemize}


\section{Alignment of States after Symmetry Breaking} 

As stated, to first approximation all 24 tetrahedron states are identical 
(symmetric limit). Although the symmetry breaking that makes them 
different can have various origins, it must have to do with a 
breaking of the new, superstrong interaction that keeps the 
4 tetrons together. 

In more concrete terms one may ask, what properties the 
tetrons need to form the 24 states and in particular to make them all 
different. 
I have considered various possibilities but will present here only the 
most appealing: namely, assuming non-identical tetrons, I demand 
that these appear in 4 different 'charge states' called 
$\chi(q_i,x)$, $i=1,2,3,4$ fulfilling the  
following selection rules: In a tetrahedron bound state with 4 tetrons 
\begin{itemize}
\item (A) each charge $q_i$ which can take one of 4 possible values 
appears once and only once 
\item (B) the sum of charges vanishes, i.e. the bound 
states are singlets under the new superstrong interaction. 
\end{itemize}
From (A) we directly get 24 product states    
\begin{equation} 
\phi_\sigma (x_1,x_2,x_3,x_4)=
\chi(q_{\sigma(1)},x_1) \chi(q_{\sigma(2)},x_2)
\chi(q_{\sigma(3)},x_3) \chi(q_{\sigma(4)},x_4)   
\label{eq44bb}
\end{equation}
corresponding to the 24 permutations $\sigma \in S_4$ 
and where I have used the notation 
$\sigma:(1,2,3,4)\rightarrow (\sigma(1),\sigma(2),\sigma(3),\sigma(4))$.
Problem: At this point the 24 states can still be transformed 
into each other by a suitable rotation of the tetrahedron (permutation 
of the $x_i$). In order to get them different I demand additionally 
\begin{itemize}
\item (C) the existence of cores or "nuclei" with 4 different 
charges $Q_i$ in the centre of each tetron $i$, i.e. in the corners 
of the tetrahedron, which are surrounded by the charges 
$q_{\sigma(i)}$. The range of possible values of the $Q_i$ 
may be chosen identical to that of the $q_i$. 
\end{itemize}
The wave function for the tetrahedrons then reads 
\begin{equation} 
\phi_\sigma (x_1,x_2,x_3,x_4)=
\chi(q_{\sigma(1)},Q_1,x_1) \chi(q_{\sigma(2)},Q_2,x_2)
\chi(q_{\sigma(3)},Q_3,x_3) \chi(q_{\sigma(4)},Q_4,x_4)   
\label{eq44}
\end{equation}
and, modulo rotations, there are now 24 different tetrahedron states 
indexed by $\sigma \in S_4$. 
Equivalently one may say, that a tetron is described by 2 quantum 
numbers q and Q fulfilling the above restrictions and selection rules. 

Note that I have not yet specified the nature of these quantum numbers. 
At the moment they are just properties which distinguish the tetrons. 
Furthermore, in order that eq. (\ref{eq44}) holds, 
one is not really tied to the geometrical model of the 
tetrahedron. It would suffice to have objects which can be 
combined in all different ways with respect to 2 properties q and Q. 
It is just for definiteness, that I am calling the 4-tetron 
bound states 'tetrahedrons'. 

Using $\sigma(1)=a$, $\sigma(1)=b$, $\sigma(1)=c$ and $\sigma(1)=d$ 
one may rewrite the last equation in various forms: 
\begin{equation} 
\phi_{\sigma}=|abcd>=\phi_{\overline{abcd}}=\chi (q_a,Q_1) \otimes \chi(q_b,Q_2) 
\otimes \chi (q_c,Q_3) \otimes \chi (q_d,Q_4)   
\label{eq44a}
\end{equation}
where I have used the notation 
$\sigma=\overline{abcd}: 1234 \rightarrow \overline{abcd}$.

{\footnotesize
In some sense the tetrahedron is similar to a chemical molecule 
with nuclei $Q_i$ and wave function clouds $q_j$. 
Conditions (A) and (B) ensure that the lowest lying orbitals 
cannot be infinitely filled. 

To understand the breaking more clearly, consider the simplified case of 
2 clouds $q_+$ and $q_-$ surrounding 2 cores $Q_+$ and $Q_-$ and 
forming 2-core-2-cloud bound states, namely 
\begin{equation} 
\phi_{+-}= \chi (q_+,Q_+)  \otimes \chi(q_-,Q_-)
\label{eq44f}
\end{equation}
and
\begin{equation} 
\phi_{-+}= \chi (q_-,Q_+)  \otimes \chi(q_+,Q_-)
\label{eq44g}
\end{equation}
and no others (i.e. assuming a modified selection rule that only bound 
states with 2 different clouds and 2 different cores exist). 
For state $\phi_{+-}$ cloud $q_+$ is nearer to core $Q_+$ whereas 
in $\phi_{-+}$ it is nearer to $q_-$, and analogously for cloud 
$q_-$. Because of all charges being different (i.e. due to the 
breaking of permutation symmetry) the states $\phi_{+-}$ 
and $\phi_{-+}$ will not be degenerate. One of them (representing 
the neutrino) will be lower in mass than the other (representing 
the electron). In nonrelativistic perturbation theory one 
would generically have mass formulas 
\begin{eqnarray} 
E_{+-}&=& E(q_+,Q_+)+ E(q_-,Q_-) \nonumber \\
&+&<\phi_{+-}|V(q_+,q_-)+V(Q_+,Q_-)+V(q_+,Q_-)+V(q_-,Q_+)|\phi_{+-}>
\label{eq44h}
\end{eqnarray} 
\begin{eqnarray}  
E_{-+}&=& E(q_-,Q_+)+ E((q_+,Q_-) \nonumber \\
&+&<\phi_{-+}|V(q_+,q_-)+V(Q_+,Q_-)+V(q_-,Q_-)+V(q_+,Q_+)|\phi_{-+}>
\label{eq44i}
\end{eqnarray} 
where $V(a,b)$ denotes the interaction between charge a and b and 
$E(a,b)$ denotes the lowest order energy eigenvalues of completely 
separated tetrons. 
Mass generation by breaking terms of the new interactions 
will be further addressed in the section 4. 
}

Note that even though most of the time I consider tetrahedrons 
at rest, the framework eq. (\ref{eq44a}) is suitable general to be 
applicable to relativistic tetrahedron states by considering spacetime 
instead of space coordinates. 
Then, if one dislikes the geometrical notion of 'cores' and 'clouds', 
one may, on a somewhat more abstract level, consider the 
$\phi_{\overline{abcd}}$ to be products of 
16 complex fields $\chi_i^a$, $i,a=1,2,3,4$ 
\begin{equation} 
\phi_{\overline{abcd}}=\chi_{1}^{a} \chi_{2}^{b} \chi_{3}^{c} \chi_{4}^{d} 
\label{eq44ahh22}
\end{equation}
at spacetime point x. Compared to eq. (\ref{eq44a}) the role of the 
charges Q (q) is played by the lower (upper) indices. 
If one looks at this equation, there is one question immediately 
arising: how can the products of tetrons $\chi$ on the right hand side 
become Dirac spinors $\phi_{\overline{abcd}}$? An answer to this 
question will be given in a forthcoming paper \cite{lampxx}. 

Next, it must be realized 
that the physical states are linear combinations of the 
product states eq. (\ref{eq44}). Consider, for example, the 4 states 
generated by applying the subgroup $Z_4$ on 
the unit element of $S_4$, i.e. $|1234>$, $|2341>$, $|3412>$ and $|4123>$. 
They are naturally built into a singlet  
\begin{equation} 
\phi_{\nu_\tau} =\frac{1}{\sqrt{4}}[\phi_{\overline{1234}} 
+ \phi_{\overline{2341}} + \phi_{\overline{3412}} + \phi_{\overline{4123}}] 
\label{eq45}
\end{equation}
representing the $\tau$-neutrino and 3 nonsinglet combinations 
\begin{eqnarray} 
\phi_{t_1} &=&\frac{1}{\sqrt{4}}[\phi_{\overline{1234}} + i \phi_{\overline{2341}} - 
\phi_{\overline{3412}} -i \phi_{\overline{4123}}]   \\
\phi_{t_2} &=&\frac{1}{\sqrt{4}}[\phi_{\overline{1234}} - \phi_{\overline{2341}} + 
\phi_{\overline{3412}} - \phi_{\overline{4123}}]  \\
\phi_{t_3} &=&\frac{1}{\sqrt{4}}[\phi_{\overline{1234}} -i \phi_{\overline{2341}} - 
\phi_{\overline{3412}} +i \phi_{\overline{4123}}]
\label{eq46}
\end{eqnarray}
which are degenerate in energy, as shown in the next section, 
and will be used to represent the 3 color states 
of the top-quark. 

The reason for considering these linear combinations instead of 
the simple product state eq. (\ref{eq44}) is that they turn out to be 
eigenfunctions of the $U(1)_{B-L}\times SU(3)_c$ charge operators 
$\lambda_3$, $\lambda_8$ and $Y_{B-L}$. 

To prove this, one first shows that the states (\ref{eq45})-(\ref{eq46}) 
are eigenfunctions of permutation operators $R_0=\overline{1234}$,  
$R_1=\overline{2341}$, $R_2=\overline{3412}$ and $R_3=\overline{4123}$ 
and afterwards constructs the $U(1)_{B-L}\times SU(3)_c$ charges as 
linear combinations of the $R_i$. 
In fact, writing $\phi_{\nu_\tau}=(1,0,0,0)$, $\phi_{t_1}=(0,1,0,0)$ etc, 
the action of the $R_j$ on the states (\ref{eq45})-(\ref{eq46}) 
is given by $R_0=(1,1,1,1)$, $R_1=diag(1,-i,-1,i)$, 
$R_2=diag(1,-1,1,-1)$ and $R_3=diag(1,i,-1,-i)$. 
Therefore, one has 
$Y_{B-L}=-\frac{1}{6}(3,-1,-1,-1)=-\frac{1}{6}(R_1+R_2+R_3)$ 
and similarly for $\lambda_3$ and $\lambda_8$.  

Of course, at this point $Z_4$ is merely a discrete 
symmetry. From the Casimirs $\lambda_3$, $\lambda_8$ and $Y_{B-L}$ 
it should be completed to a global $U(1)_{B-L}\times SU(3)_c$ 
and afterwards in the limit when the tetrahedron shrinks to a point-like 
fermion an effective local gauge symmetry should be constructed. 

Note that with the help of suitable permutations the above 
analysis can be extended to any other quark-lepton quadruplet. 
Note further that using $SU(4)$ instead of $U(1)_{B-L}\times SU(3)_c$ 
one would in general get leptoquarks besides gluon interactions and the 
$U(1)_{B-L}$-photon. These, however, can be shown to be forbidden 
in the present model. I shall come back to this point in 
section 5. 

If one prefers to start with the Kleinsche Vierergruppe $K$ instead of 
$Z_4$ one has an alternative representation for $\phi_{\nu_\tau}$ and 
$\phi_{t_i}$ 
\begin{eqnarray} 
\phi_{\nu_\tau} &=&\frac{1}{\sqrt{4}}[\phi_{\overline{1234}} 
+ \phi_{\overline{2143}} + \phi_{\overline{3412}} + \phi_{\overline{4321}}] 
\label{eq46700} \\
\phi_{t_1} &=&\frac{1}{\sqrt{4}}[\phi_{\overline{1234}} + \phi_{\overline{2143}} - 
\phi_{\overline{3412}} -\phi_{\overline{4321}}]   \\
\phi_{t_2} &=&\frac{1}{\sqrt{4}}[\phi_{\overline{1234}} - \phi_{\overline{2143}} + 
\phi_{\overline{3412}} -\phi_{\overline{4321}}]  \\
\phi_{t_3} &=&\frac{1}{\sqrt{4}}[\phi_{\overline{1234}} - \phi_{\overline{2143}} - 
\phi_{\overline{3412}} +\phi_{\overline{4321}}]
\label{eq4670}
\end{eqnarray}
which are eigenstates of the permutation operators 
$K_0=\overline{1234}=(1,1,1,1)$, $K_1=\overline{2143}=(1,1,-1,-1)$, 
$K_2=\overline{3412}=(1,-1,1,-1)$ and $K_3=\overline{4321}=(1,-1,-1,1)$ 
and correspondingly a different representation for the
Casimirs $\lambda_3=\frac{1}{2}(K_1-K_2)$, 
$\lambda_8=\frac{1}{2\sqrt{3}} (K_1+K_2-2K_3)$ and 
$Y_{B-L}=-\frac{1}{6}(3,-1,-1,-1)=-\frac{1}{6}(K_1+K_2+K_3)$. 
We shall see shortly whether $K$ oder $Z_4$ 
is preferable. 

Having dealt with $Z_4$ (or $K$), one can treat the rest of 
$S_4= Z_4 \diamond Z_3 \diamond Z_2$ in a similar fashion. 
Namely, for $S_2=Z_2$ one may define 2 states 
\begin{equation} 
\phi_{\pm}=\frac{1}{\sqrt{2}}[\phi_{\overline{13}}\pm\phi_{\overline{31}}] 
\label{eq4833}
\end{equation}
corresponding to eqs. (\ref{eq44f}) and (\ref{eq44g}) 
which are eigenstates of the generator $\overline{31}$ of $S_2$ 
with eigenvalue +1 and -1, i.e. these states should be 
identified with the two partners of a weak isospin 
doublet (like the electron and its neutrino). 
\footnote{I consider $S_2$ to be the permutations 
of two objects called here 1 and 3 to be in agreement with table 1.}
 
One can easily construct a set of Pauli matrices $\sigma_i$, $i=1,2,3$ 
for the states eq. (\ref{eq4833}) by using $\overline{31}$ as $\sigma_3$ 
and then defining creation and annihilation 
operators $\sigma_+\phi_-=\phi_+$ and $\sigma_-\phi_+=\phi_-$ 
and from these $\sigma_1=\sigma_+ + \sigma_-$ 
and $\sigma_2=i(\sigma_+ - \sigma_-)$. This set of 
matrices is easily seen to obey SU(2) commutation 
relations and is going to generate the weak $SU(2)$-symmetry. 

Combining eqs. (\ref{eq46}) and (\ref{eq4833}) the lepton and 
quark states get 4 additional terms: neutrinos and 
up-type quarks with a positive, electrons and down-type 
quarks with a negative sign: 
\begin{eqnarray} 
\phi_{N/E} &=& \frac{1}{\sqrt{8}} (\phi^{F}_0 \pm \phi^{B}_0) 
\label{eq48aa} \\
\phi^{F}_0&=&\phi_{\overline{1234}}+\phi_{\overline{2143}}
          +\phi_{\overline{3412}}+\phi_{\overline{4321}} \nonumber \\
\phi^{B}_0&=&\phi_{\overline{3214}}+\phi_{\overline{4123}}
          +\phi_{\overline{1432}}+\phi_{\overline{2341}} \nonumber 
\end{eqnarray}
\begin{eqnarray} 
\phi_{U_1/D_1} &=& \frac{1}{\sqrt{4}} (\phi^{F}_1 \pm \phi^{B}_1) \\
\phi^{F}_1&=&\phi_{\overline{1234}}-\phi_{\overline{3412}} \nonumber \\
\phi^{B}_1&=&\phi_{\overline{3214}}-\phi_{\overline{1432}} \nonumber 
\end{eqnarray}
\begin{eqnarray} 
\phi_{U_2/D_2} &=& \frac{1}{\sqrt{8}} (\phi^{F}_2 \pm \phi^{B}_2) \\
\phi^{F}_2&=&\phi_{\overline{1234}}-\phi_{\overline{2143}}
          +\phi_{\overline{3412}}-\phi_{\overline{4321}} \nonumber \\
\phi^{B}_2&=& \phi_{\overline{3214}}-\phi_{\overline{4123}}
          +\phi_{\overline{1432}}-\phi_{\overline{2341}} \nonumber 
\end{eqnarray}
\begin{eqnarray} 
\phi_{U_3/D_3} &=& \frac{1}{\sqrt{4}} (\phi^{F}_3 \pm \phi^{B}_3) 
\label{eq48} \\
\phi^{F}_3&=&\phi_{\overline{4321}}-\phi_{\overline{2143}} \nonumber \\
\phi^{B}_3&=&\phi_{\overline{2341}}-\phi_{\overline{4123}} \nonumber 
\end{eqnarray}
In the language of molecular physics these are 
the symmetry adapted wave function of the dihedral group $D_4$.
$D_4$ is of order 8. It contains the even elements of the Kleinsche 
Vierergruppe combined with the odd permutation 
$\overline{3214}=(1\leftrightarrow 3)$ and according to table 1 
generates the first family. 

I have made the assignments in such a way that isospin partners 
are obtained by changing the sign of the odd permutation functions. 
This is in accord with eq. (\ref{eq4833}) 
but if we do so in the general case, the $Z_4$-functions 
eqs. (\ref{eq45})-(\ref{eq46}) are 
out of the game, because $\phi_{t_1}$ and $\phi_{t_3}$ 
would have to be isospin partners. Nevertheless, it is 
possible to save the $Z_4$-scenario by choosing a somewhat 
different assignment in eqs. (\ref{eq48aa})-(\ref{eq48})
and then demanding that the germ of weak $SU(2)$ does not actually 
correspond to the permutation $\overline{3214}=(1\leftrightarrow 3)$ 
as in eq. (\ref{eq4833}) but to the transitions  
$\phi_N\leftrightarrow \phi_E$ and 
$\phi_{U_i}\leftrightarrow \phi_{D_i}$.
Since this is somewhat artificial I think that the 
$K$ is the better option. 

Finally there is a state-mixing due to the family group $Z_3 \subset S_3$ 
which means that instead of ...1234123...+backward, ...2134...+backward and 
...4231...+backward of section 2 the 3 families correspond to 
\begin{eqnarray} 
\phi_{A}&=&\frac{1}{\sqrt{3}}[\phi_{id}+\phi_{g_1}+\phi_{g_2}] \nonumber \\
\phi_{B}&=&\frac{1}{\sqrt{3}}[\phi_{id}+\alpha\phi_{g_1}+\alpha^2\phi_{g_2}] \nonumber \\
\phi_{C}&=&\frac{1}{\sqrt{3}}[\phi_{id}+\alpha^2\phi_{g_1}+\alpha\phi_{g_2}] 
\label{eq49}
\end{eqnarray}
where $\alpha=exp(2i\pi /3)$ and $id$, $g_1$ and $g_2$ denote the 
elements of $Z_3$, i.e. even permutations of $S_3$, $id=\overline{123}$
$g_1=\overline{231}$ and $g_1=\overline{312}$. In the notation of eq. 
(\ref{eq44}) one has 
$\phi_{\sigma}=\chi(q_{\sigma(1)},Q_1)\otimes \chi(q_{\sigma(2)},Q_2)
\otimes \chi(q_{\sigma(3)},Q_3)$ for $\sigma \in S_3$, e.g. 
$\phi_{g_1}=\chi(q_2,Q_1)\otimes \chi(q_3,Q_2)\otimes \chi(q_1,Q_3)$. 

The reason to form the combinations eq. (\ref{eq49})-(\ref{eq49}) 
is that they are idempotent and orthogonal eigenstates of the "charge 
operators" $g_1$ and $g_2$, because of the following relations
\begin{eqnarray} 
g_1 \phi_{A} &=& \phi_{A}             \nonumber \\
g_1 \phi_{B} &=& \alpha^2\phi_{B}      \nonumber \\
g_1 \phi_{C} &=& \alpha\phi_{C}        \nonumber \\
g_2 \phi_{A} &=& \phi_{A}              \nonumber \\
g_2 \phi_{B} &=& \alpha\phi_{B}        \nonumber \\
g_2 \phi_{C} &=& \alpha^2\phi_{C}       
\label{eq492}
\end{eqnarray}
which follow easily from the properties $g_1^2=g_2$, $g_1 g_2=id$ 
and $g_2^2=g_1$.  

Extending this to $S_3$ one gets the 6 expressions 
\begin{eqnarray} 
\phi_{\nu_{\tau}}&=&\frac{1}{\sqrt{6}}
 [\phi_{id}+\phi_{g_1}+\phi_{g_2}+\phi_{u_0}+\phi_{u_1}+\phi_{u_2}] \nonumber \\
\phi_{\tau}&=&\frac{1}{\sqrt{6}}
 [\phi_{id}+\phi_{g_1}+\phi_{g_2}-\phi_{u_0}-\phi_{u_1}-\phi_{u_2}] \nonumber \\
\phi_{\nu_{\mu}}&=&\frac{1}{\sqrt{12}}
 [2\phi_{id}-\phi_{g_1}-\phi_{g_2}+2\phi_{u_0}-\phi_{u_1}-\phi_{u_2}] \nonumber \\
\phi_{\mu}&=&\frac{1}{\sqrt{12}}
 [2\phi_{id}-\phi_{g_1}-\phi_{g_2}-2\phi_{u_0}+\phi_{u_1}+\phi_{u_2}] \nonumber \\
\phi_{\nu_{e}}&=&\frac{1}{2}
 [\phi_{g_1}-\phi_{g_2}  -\phi_{u_1}+\phi_{u_2}] \nonumber \\
\phi_{e}&=&\frac{1}{2}
 [-\phi_{g_1}+\phi_{g_2}-\phi_{u_1}+\phi_{u_2}] 
\label{eq49aaa}
\end{eqnarray}
for the leptons.

With these building blocks in mind one may write down the complete formula 
for each fermion as a specific linear combination of the 24 product states 
eq. (\ref{eq44a}): 
\begin{equation} 
\phi^J=\sum_{a,b,c,d=1}^{24} \lambda_{abcd}^J |abcd>
\label{eq500}
\end{equation}
where $J=\nu_e, e, ...$ numbers the 24 elementary fermions. 
The representations $A_1$, $A_2$ and $E$ of $S_4$ with dimension 1, 1 
and 2 respectively are reserved for leptons, whereas the 3-dimensional 
representations $T_1$ and $T_2$ are used for up- and down-type quarks, 
respectively (for more details of the $S_4$-representations see the 
appendix). 
The $\tau$-neutrino, for example, then corresponds to the overall singlet 
\begin{eqnarray} 
\phi_{\nu_\tau}=\frac{1}{\sqrt{24}}\{
\phi_{\overline{1234}} + \phi_{\overline{2143}} + 
\phi_{\overline{3412}} + \phi_{\overline{4321}}
+ \phi_{\overline{3214}} + \phi_{\overline{2341}}  
+ \phi_{\overline{1432}} + \phi_{\overline{4123}}              \nonumber \\
+\phi_{\overline{2134}} + \phi_{\overline{1243}} + 
\phi_{\overline{3421}} + \phi_{\overline{4312}}
+ \phi_{\overline{3124}} + \phi_{\overline{1342}}  
+ \phi_{\overline{2431}} + \phi_{\overline{4213}}               \nonumber \\
+\phi_{\overline{4231}} + \phi_{\overline{2413}} + 
\phi_{\overline{3142}} + \phi_{\overline{1324}}
+ \phi_{\overline{3241}} + \phi_{\overline{2314}}  
+ \phi_{\overline{4132}} + \phi_{\overline{1423}}           \}
\label{eq5088}
\end{eqnarray}
and the wave function for the $\tau$-lepton is given by 
\begin{eqnarray} 
\phi_\tau=\frac{1}{\sqrt{24}}\{
\phi_{\overline{1234}} + \phi_{\overline{2143}} + 
\phi_{\overline{3412}} + \phi_{\overline{4321}}
- \phi_{\overline{3214}} - \phi_{\overline{2341}} - 
\phi_{\overline{1432}} - \phi_{\overline{4123}}              \nonumber \\
-\phi_{\overline{2134}} - \phi_{\overline{1243}} - 
\phi_{\overline{3421}} - \phi_{\overline{4312}}
+ \phi_{\overline{3124}} + \phi_{\overline{1342}} + 
\phi_{\overline{2431}} + \phi_{\overline{4213}}               \nonumber \\
-\phi_{\overline{4231}} - \phi_{\overline{2413}} - 
\phi_{\overline{3142}} - \phi_{\overline{1324}}
+ \phi_{\overline{3241}} + \phi_{\overline{2314}} + 
\phi_{\overline{4132}} + \phi_{\overline{1423}}           \}
\label{eq5xy20}
\end{eqnarray}
whereas the second color component of the strange quark reads 
\begin{eqnarray} 
\phi_{s_2}=\frac{1}{\sqrt{8}}\{
\phi_{\overline{1234}} - \phi_{\overline{2143}} + 
\phi_{\overline{3412}} - \phi_{\overline{4321}}
+ \phi_{\overline{1432}} + \phi_{\overline{2341}}  
- \phi_{\overline{2134}} - \phi_{\overline{4312}}           \}
\label{eq50xx3}
\end{eqnarray}
This is obtained from the 22 matrix element of $T_1$. 
Its weak isospin partner $\phi_{c_2}$ (= the 22 matrix 
element of $T_2$) is obtained by changing 
the signs of the odd permutations in eq. (\ref{eq50xx3}). 

\section{A phenomenological Approach to Masses and Charges}

As stated, to first approximation all 24 states are identical 
(symmetric limit). In this limit all masses are equal. 
Since the symmetry breaking mechanism must have to do with 
the (new) interaction that keeps the 4 tetrons together, 
the interaction must have a $S_4$-breaking part $H_X$ 
and (using for simplicity nonrelativistic framework) in first order 
perturbation theory the masses would be calculable as 
\begin{equation} 
m_{\overline{abcd}}=<abcd|H_X|abcd>
\label{eq5522}
\end{equation}

One could then use the charge eigenstates eq. (\ref{eq500})
to calculate the matrix elements eq. (\ref{eq5522}). If one further assumes that 
\begin{equation} 
H_X=V(x_1,x_2,x_3,x_4)=V_{12}+V_{13}+V_{14}+V_{23}+V_{24}+V_{34}
\label{eq711}
\end{equation}
with $V_{ij}=q_i q_j V_0(|x_i - x_j|)$, one can easily calculate 
\begin{equation} 
m_J = <\phi^J|V|\phi^J>=\sum_{abcd,a'b'c'd'}
\lambda^{\ast}_{abcd}\lambda_{a'b'c'd}<abcd|V|a'b'c'd'>
\label{eq712}
\end{equation}
with 
\begin{eqnarray} 
<abcd|V|a'b'c'd'> &=& \delta_{cc'}\delta_{dd'} q_1 q_2 
[\delta_{aa'}\delta_{bb'}J^C_{ab}+\delta_{ab'}\delta_{ba'}J^A_{ab}] \nonumber \\
& &+\delta_{bb'}\delta_{dd'} q_1 q_3
[\delta_{aa'}\delta_{cc'}J^C_{ac}+\delta_{ac'}\delta_{ca'}J^A_{ac}] \nonumber \\
& &+\delta_{bb'}\delta_{cc'}  q_1 q_4
[\delta_{aa'}\delta_{dd'}J^C_{ad}+\delta_{ad'}\delta_{da'}J^A_{ad}] \nonumber \\
& &+\delta_{aa'}\delta_{dd'}  q_2 q_3
[\delta_{bb'}\delta_{cc'}J^C_{bc}+\delta_{bc'}\delta_{cb'}J^A_{bc}] \nonumber \\
& &+\delta_{aa'}\delta_{cc'}  q_2 q_4
[\delta_{bb'}\delta_{dd'}J^C_{bd}+\delta_{bd'}\delta_{db'}J^A_{bd}] \nonumber \\
& &+\delta_{aa'}\delta_{bb'}  q_3 q_4
[\delta_{cc'}\delta_{dd'}J^C_{cd}+\delta_{cd'}\delta_{dc'}J^A_{cd}]
\label{eq714}
\end{eqnarray}

$J^A$ und $J^C$ are generalizations of exchange and Coulomb integrals 
\begin{equation} 
J^C_{ab} = \int dx^4_i d^4x_j V_0(|x_i - x_j|) \phi_a^{\ast}(x_i)
\phi_b^{\ast}(x_j) \phi_{a}^{\ast}(x_i) \phi_{b}^{\ast}(x_j)
\label{eq713}
\end{equation}
\begin{equation} 
J^C_{ab} = \int dx^4_i d^4x_j V_0(|x_i - x_j|) \phi_a^{\ast}(x_i)
\phi_b^{\ast}(x_j) \phi_{a}^{\ast}(x_i) \phi_{b}^{\ast}(x_j)
\label{eq715}
\end{equation}
They are symmetric in the interchange of a and b, so that 
besides the 4 unknowns $q_i$ one has 20 unknown integrals. 

We do not undertake to examine this further, because apart from 
the many unknown integrals there is the additional uncertainty about 
the validity of the nonrelativistic approach. 
The nonrelativistic picture with a Hamiltonian of the 
form $H=\sum_i E_{kin}^i + H_X$ with $H_X=V=\sum_{ij} V(|x_i-x_j|)$ 
is qualitatively nice to give a good overview what states exist. However, 
it is very unlikely that it works quantitatively correct. 

Instead I shall analyze the mass matrix of the fermions 
as it comes out from their description of the form of 
linear combinations of tetrahedron states eq. (\ref{eq500}). 
A complete analysis of these states would involve a 24 times 24 mass 
matrix. 
We shall not attempt this here but to get an intuition 
about what can be achieved we split the problem into 3 pieces 
\begin{itemize}
\item Mass splittings among the weak isospin dublets: in the 
present model weak isospin states like $\nu_e$ and $e$ are 
described by $S_2$ wave functions $\phi_{\pm}$ eq. (\ref{eq4833}) which are 
interaction eigenstates with mass matrix 
\begin{equation} 
m_{\pm}=\frac{1}{2}
\left(\begin{array}{cc}
1 & \pm 1 
\end{array}\right)
\left(\begin{array}{cc}
m_{11} & m_{12} \\
m_{21} & m_{22}
\end{array}\right)
\left(\begin{array}{c}
1 \\
\pm 1
\end{array}\right)
\end{equation} 
It is then easy to accomodate a mass structure $m_e >> m_{\nu_e}$ 
namely with a "democratic" mass matrix $m_{11}=m_{12}=m_{21}=m_{22}$. 

\item Mass splittings among the families: in the present model 
family states like $e$, $\mu$ and $\tau$ are described by 
$Z_3$ wave functions $\phi_{A,B,C}$ eq. (\ref{eq49}), 
which are interaction eigenstates with mass matrix  
\begin{equation} 
m_{A,B,C}=\frac{1}{3}
\left(\begin{array}{ccc}
1 & 1,\alpha^*,\alpha^{*2} & 1 ,\alpha^{*2},\alpha^* 
\end{array}\right)
\left(\begin{array}{ccc}
m_{11} & m_{12} & m_{13}  \\
m_{21} & m_{22} & m_{23}  \\
m_{31} & m_{32} & m_{33}
\end{array}\right)
\left(\begin{array}{c}
1 \\
1,\alpha,\alpha^{2} \\
1,\alpha^{2},\alpha
\end{array}\right)
\end{equation} 
Using the algebraic identity $1+\alpha +\alpha^{2} = 0$ which implies 
$Im(\alpha)=-Im(\alpha^{2})$ and assuming the mass matrix to be symmetrical, 
it is easy to show that 
\begin{eqnarray} 
m_A&=&\frac{1}{3} (tr(m)+2m_n)   \\
m_B&=&m_C=\frac{1}{3} (tr(m)-m_n) 
\label{eq71455}
\end{eqnarray}
where $m_n=m_{12}+m_{13}+m_{23}$ 
and one immediately sees that by choosing $tr(m)=m_n$ one can obtain 
a heavy family with $m_\tau=m_A$ and two light ones $m_B=m_C=0$. 

\item Mass splittings among the quarks and leptons: in the present model 
quark and lepton states $E$ and $D$ are distinguished by their 
behavior under $Z_4$ (or $K$) and therefore described 
by the wave functions $\phi_{E}$ and $\phi_{D_{1,2,3}}$, 
eqs. (\ref{eq45})-(\ref{eq46}) (or (\ref{eq46700})-(\ref{eq4670})) 
which are interaction eigenstates with mass matrix 
\begin{eqnarray}
m_{E,D_1,D_2,D_3}&=& \frac{1}{4}
\left(\begin{array}{cccc}
1,1,1,1 & 1,-i,-1,i & 1,-1,1,-1 & 1,i,-1,-i 
\end{array}\right)    \nonumber \\
& & \left(\begin{array}{cccc}
m_{11} & m_{12} & m_{13} & m_{14}  \\
m_{21} & m_{22} & m_{23} & m_{24}  \\
m_{31} & m_{32} & m_{33} & m_{34}  \\
m_{41} & m_{42} & m_{43} & m_{44}  
\end{array}\right)
\left(\begin{array}{c}
1,1,1,1 \\
1,i,-1,-i \\
1,-1,1,-1 \\
1,-i,-1,i 
\end{array}\right)
\label{eq99460}
\end{eqnarray}
The requirement $m_{D_1}=m_{D_2}=m_{D_3}$ of quark masses being 
color independent directly leads to 
\begin{eqnarray} 
m_E&=& \frac{1}{4} (tr(m)+6m_n) \\
m_{D_1}&=&m_{D_2}=m_{D_3}= \frac{1}{4} (tr(m)-2m_n) 
\label{eq714uuv}
\end{eqnarray}
where $m_n=m_{13}+m_{24}$ 
and one is this way able to accomodate any quark lepton 
mass ratio one likes. For $K$ instead of $Z_4$ as symmetry group 
one obtains a result which is formally 
identical. 
\end{itemize}

Notes added: \\
1. Although this approach knows nothing about the true nature of the 
tetron interactions and relies solely on the symmetry properties 
of the tetrahedrons, it can in principle be used to calculate all 
mass ratios of the fermion spectrum. 
The basic mass scale is of course set by the strength 
of the new interaction, but mass ratios can be 
inferred from symmetry principles. 
\\
2. Almost identical results can be obtained on the 
basis of the approach eq. (\ref{eq5522}), if one supposes that the matrix elements 
transform according to a representation of $S_4$ or $S_3$ as 
described in the appendix. This will not be discussed 
here. 

The above procedure for masses may be applied as 
well to standard model fermion charges, provided one assumes that 
each charge C corresponds to an operator $C_{op}$ acting on 
the linear combinations of states $\phi^J$ eq. (\ref{eq500}) as 
\begin{equation} 
C_{op} \phi^J = C_J \phi^J
\label{eq415}
\end{equation}
In the left-right symmetric standard model 
$U(1)_{B-L}\times SU(3)_c \times SU(2)_L \times SU(2)_R$ 
there are the following charges: 
$Y_{B-L}$ (B-L charge), $C_8$ and $C_3$ (color charges) and 
$T_{3L}$ and $T_{3R}$ for weak isospin. 

\begin{itemize}
\item Starting with weak isospin dublets and using 
that weak isospin states are described in the present model  
by $S_2$ wave functions $\phi_{\pm}$ eq. (\ref{eq4833}) with 
eigenvalues $\pm \frac{1}{2}$ we require  
\begin{equation} 
\pm \frac{1}{2} = \frac{1}{2}
\left(\begin{array}{cc}
1 & \pm 1 
\end{array}\right)
\left(\begin{array}{cc}
c_{11} & c_{12} \\
c_{21} & c_{22}
\end{array}\right)
\left(\begin{array}{c}
1 \\
\pm 1
\end{array}\right)
\end{equation} 
We therefore find that 
\begin{equation} 
T_{3op} = 
\left(\begin{array}{cc}
c_{11} & c_{12} \\
c_{21} & c_{22}
\end{array}\right) = 
\left(\begin{array}{cc}
0 & 1 \\
1 & 0
\end{array}\right)
\end{equation} 
and the analysis is identical for left and righthanded fermions. 
\item Going next to $B-L$ and to the color charges we have for the 
eigenvalues of a quark-lepton quadruplet $L$ and $Q$ in $Z_4$ the 
following condition:
\begin{eqnarray}
C(L,Q_1,Q_2,Q_3)&=& \frac{1}{4}
\left(\begin{array}{cccc}
1,1,1,1 & 1,-i,-1,i & 1,-1,1,-1 & 1,i,-1,-i 
\end{array}\right)  \nonumber    \\
& & \left(\begin{array}{cccc}
c_{11} & c_{12} & c_{13} & c_{14}  \\
c_{21} & c_{22} & c_{23} & c_{24}  \\
c_{31} & c_{32} & c_{33} & c_{34}  \\
c_{41} & c_{42} & c_{43} & c_{44}  
\end{array}\right)
\left(\begin{array}{c}
1,1,1,1 \\
1,i,-1,-i \\
1,-1,1,-1 \\
1,-i,-1,i 
\end{array}\right)
\end{eqnarray}
where on the left hand side there are the standard model values 
$C_8(L)=0$, $C_8(Q_1)=1$, $C_8(Q_2)=-2$, $C_8(Q_3)=1$ and 
similarly $C_3(L)=0$, $C_3(Q_1)=1$, $C_3(Q_2)=0$, $C_3(Q_3)=-1$ 
and $Y_{B-L}(L)=-\frac{1}{2}$, $Y_{B-L}(Q_1)=Y_{B-L}(Q_2)=Y_{B-L}(Q_3)=\frac{1}{6}$. 
With a little algebra one obtains 
\begin{eqnarray} 
Y_{B-L}&=&-\frac{1}{3} 
\left(\begin{array}{cccc}
0 & 1 & 1 & 1  \\
1 & 0 & 1 & 1  \\
1 & 1 & 0 & 1  \\
1 & 1 & 1 & 0  
\end{array}\right) 
\label{eq7145595} \\
C_3&=&\frac{1}{2} 
\left(\begin{array}{cccc}
0 & 1 & 0 & 1  \\
-1 & 0 & 1 & 0  \\
0 & -1 & 0 & 1  \\
-1 & 0 & -1 & 0  
\end{array}\right) \\
\label{eq7145593}
C_8&=&\frac{1}{2}
\left(\begin{array}{cccc}
0 & 1 & -2 & 1  \\
1 & 0 & 1 & -2  \\
-2 & 1 & 0 & 1  \\
1 & -2 & 1 & 0  
\end{array}\right)
\label{eq7145591}
\end{eqnarray}
Using $K$ instead of $Z_4$ the same form is obtained for $Y_{B-L}$ 
and $C_8$ but $C_3$ is modified to read: 
\begin{eqnarray} 
C_3&=&\frac{1}{2} 
\left(\begin{array}{cccc}
0 & 1 & 0 & -1  \\
1 & 0 & -1 & 0  \\
0 & -1 & 0 & 1  \\
-1 & 0 & 1 & 0  
\end{array}\right) 
\label{eq9209}
\end{eqnarray}
\end{itemize}
These charge operators can be extended to 24 times 
24 matrices acting on the fermion wave functions eq. (\ref{eq500}). 
They cannot be traced to charges of a single tetron, but 
correspond to properties of the tetrahedron wave 
functions $\phi_\sigma$. As will be shown, there is a direct 
relation to the construction of vector boson states in the 
next section. 

\section{The Model for Vector Bosons}

Vector Bosons are bound states of 8 tetrons which arise 
when 2 of the fermion-tetrahedrons $\phi_\sigma=\phi_{\overline{abcd}}$ 
and $\phi_{\sigma'}=\phi_{\overline{a'b'c'd'}}$ of 
eq. (\ref{eq44a}) approach each other and a and a', b and b', c and c' 
and d and d' interact in such a way that a cube is formed. 
More precisely, I am talking 
about a tetrahedron and an anti-tetrahedron, where a and a', b and b', c and c' 
and d and d' sit on opposite corners of the cube. Note that 
we are dealing with Spin 1 objects here because of the spin $\frac{1}{2}$ 
nature of the tetrahedrons discussed in section 2. 
When forming fermion-antifermion products one should therefore 
in principle write 
$\bar \phi_{\sigma'} \gamma_\mu \phi_\sigma$. 
To keep things manageable, however, the Dirac structure will 
not be made explicit in the formulas below. Instead I shall simply 
write $\phi^*_{\sigma'} \phi_\sigma$. 
Also, as stressed before, the geometrical picture of tetrahedrons 
and cubes should not be taken too literally. It is just a memo for 
the behavior of the permutation group $S_4$. 

In the symmetric limit, in which all tetrons are identical, there are 
24 $\times$ 24 identical cube states. When the symmetry is 
broken with charges $q_a$ ... $q_{d'}$, $Q_a$ ... $Q_{d'}$, one 
has 24$^2$ different states generated by applying the elements 
of $S'_4 \times S_4$ on $\phi^*_{\sigma'} \phi_\sigma$. One is 
thus confronted with a lot of different tetrahedron-antitetrahedron bound states 
which mediate a lot of interactions like inter family 
interactions, leptoquarks and so on, which one does not want. 

In order to reduce the number of vector boson states there must be a 
mechanism which in the process of vector boson formation from fermions 
(i.e. the tetrahedron and anti-tetrahedron approaching each other) restores 
parts of the symmetry like, for example, the $Z_3$-(family)-part, so 
that there are no interaction particles associated with $Z_3$. Instead, $Z_3$ 
must become a symmetry of the cube which transforms e.g. the $W^+$ arising 
from $\tau^+ + \nu_\tau$ to a $W^+$ decaying into $e^+ + \nu_e$.  
A similar rule holds for the transformations 
between quarks and leptons forbidding the existence of 
leptoquarks, i.e. lepton number violating processes, 
and at the same time allowing $W^+$-mediated transitions 
e.g. between $\bar b + t$ and $\tau^+ + \nu_\tau$. 

The background behind this is a nonexistence 

THEOREM: The following statements are equivalent: 
\begin{itemize}
\item the weak interactions are universal
\item leptoquark and interfamily interactions do not exist
\end{itemize}

This theorem seems quite obvious on an abstract level and 
in particular in the present model because one needs 
interfamily and leptoquark transitions as symmetries 
and not as interactions. 
It should be noted however that in principle the symmetries 
of vector bosons under tetron transformations could be 
larger than given by the theorem. 

Later I am going to explicitly identify which transitions are 
symmetries for vector bosons and which ones are not.  
This will be done by considering step by step the 
possible fermion-antifermion states and comparing them 
to the observed vector bosons. 

{\footnotesize
Before doing this, let me discuss in some detail, what might be the 
cause of the increased symmetry, i.e. the universality 
of the weak interaction, on the tetron level. One may 
speculate that it is caused by a partwise annihilation 
\footnote{One may even consider the possibility of the 
annihilation being complete. In that case one is 
left with 24 products $\phi_{id}^* \phi_{\sigma}$ 
to form vector boson states. Due to 
$\phi_{id}^* \phi_{\sigma}=[\phi_{\sigma}^* \phi_{id}]^*
=[\phi_{id}^* \phi_{\sigma^{-1}}]^*$ this 
corresponds to 24 real degrees of freedom.}
of tetron charge clouds $q_i$ so that some corners of the cube become 
indistinguishable and the cube symmetrical under $Z_3$ 
and leptoquark transformations. As soon as this happens, these 
symmetries can be interpreted as ordinary rotation symmetries in space
(e.g. for $Z_3$: $\pm 2\pi/3$ rotations about the 3 body diagonals 
of the cube). This annihilation may affect, for example, the 'front' 
triangle of the tetrahedron which is approaching a corresponding 'front' of 
an anti-tetrahedron. Stripped off of 3 of their clouds, e.g. $q_a$, $q_b$ 
and $q_d$ the tetrahedron will then look like 
$\chi (Q_1) \otimes \chi(Q_2)\otimes \chi (q_c,Q_3) \otimes \chi (Q_4)$
and similarly for the anti-tetrahedron and the resulting cube will be 
symmetric under an $S_3$ and in particular under interchanges 
$(Q_1\leftrightarrow Q_2)$ and $(Q_1\leftrightarrow Q_4)$ 
which according to table 1 transform the families into each other. 
Note however, that this picture is at best heuristic as was the 
assignment of particle states in table 1. The point is that 
according to sect. 3 fermion states are linear combinations of 
permutation states and this complicates the situation. 
Therefore, instead of giving handwaving arguments I will 
now consider explicitly the possible fermion-antifermion states. 
}

I start with the electroweak sector and want to 
represent $W^{\pm}$ and $Z$ by suitable combinations 
$\phi^*_{\overline{a'b'c'd'}} \phi_{\overline{abcd}}$. 
A simple formula can be obtained from the wave 
functions eq. (\ref{eq4833}):  
\begin{eqnarray} 
W^-&=& ' \bar\nu_l \gamma_\mu l ' = \phi_+^* \phi_-      \nonumber \\
&=& [\chi^* (q'_1,Q'_1) \otimes \chi^* (q'_3,Q'_3) 
      + \chi^* (q'_1,Q'_3) \otimes \chi^* (q'_3,Q'_1)] \nonumber \\
& & \otimes [\chi (q_1,Q_1) \otimes \chi (q_3,Q_3) 
               - \chi (q_1,Q_3) \otimes \chi (q_3,Q_1)]
\label{eqw5}
\end{eqnarray}
and
\begin{eqnarray} 
Z&=&' \bar\nu_l \gamma_\mu \nu_l - \bar l \gamma_\mu l ' 
= |\phi_+|^2 - |\phi_-|^2 
\label{eqz5}
\end{eqnarray}
where by $Z$ I actually do not mean the physical $Z$-boson but 
what is usually called $W_3$. I will from now on leave out the primes 
in the arguments implicitly understanding that whereever a wave 
functions with an asterics appears it should get primed arguments. 
Then, under very moderate assumptions on the 
product $\otimes$, the above formulae may be simplified to  
\begin{eqnarray} 
W^\pm&=&|\chi (q_1,Q_1) \chi (q_3,Q_3)|^2 - |\chi (q_1,Q_3) \chi (q_3,Q_1)|^2 \nonumber \\
 & \pm & [\chi^* (q_1,Q_1)  \chi^* (q_3,Q_3) \chi (q_1,Q_3) \chi (q_3,Q_1) - c.c.]  \\
Z&=&2[\chi^* (q_1,Q_1)  \chi^* (q_3,Q_3) \chi (q_1,Q_3) \chi (q_3,Q_1) + c.c.] 
\label{eqz66}
\end{eqnarray}

These results can easily be generalized to make the universality 
of $W^\pm$ and $Z$ on the family level explicit.  
Namely, using the representation (\ref{eqz66}) of the wave functions 
and the $Z_3$-(family)-symmetry of the weak interactions implies 
\begin{eqnarray} 
W^-&=& \{ |\phi_{id}|^2 - |\phi_{u_0}|^2
+\phi^*_{id} \phi_{u_0} - \phi^*_{u_0}\phi_{id} \} \nonumber \\
&=& \{ |\phi_{g_1}|^2 - |\phi_{u_1}|^2
+\phi^*_{g_1} \phi_{u_1} - \phi^*_{u_1}\phi_{g_1} \} \nonumber \\
&=& \{ |\phi_{g_2}|^2 - |\phi_{u_2}|^2
+\phi^*_{g_2} \phi_{u_2} - \phi^*_{u_2}\phi_{g_2} \}  \\
Z&=& 2[ \phi^*_{id} \phi_{u_0} + \phi^*_{u_0} \phi_{id}] \nonumber   \\
&=& 2[ \phi^*_{g_1} \phi_{u_1} + \phi^*_{u_1} \phi_{g_1}] \nonumber   \\
&=& 2[ \phi^*_{g_2} \phi_{u_2} + \phi^*_{u_2} \phi_{g_2}] 
\label{eqzxx2}
\end{eqnarray}
on the basis of eqs. (\ref{eq49aaa}) for $S_3$. Furthermore, 
the validity of the nonexistence theorem, i.e. absence of 
interfamily interactions, can be explicitly verified by 
showing that terms like $\bar\nu_e \mu$, $\bar\nu_\mu \tau$  and 
$\bar\nu_\tau e$ which represent the $S_3$-symmetric form of 
interfamily interactions identically vanish. 
A possible generation mixing may also be incorporated. 

Note that instead of eq. (\ref{eq49aaa}) one should of 
course use the full $S_4$-combinations eq. 
(\ref{eq500}) as factors to be used in the representation of vector 
bosons. In this case the restrictions from the 
nonexistence theorem become more complicated in shape, but the 
principle does not change. 

In a similar fashion as eq. (\ref{eqz5}) one can use the 
representation of leptons and quarks eqs. (\ref{eq45})-(\ref{eq46}) 
for $Z_4$ or eq. (\ref{eq46700})-(\ref{eq4670}) for the Kleinsche Vierergruppe 
to write down formulae for the gluons and the photon. The photon, 
or more precisely, the $U(1)_{B-L}$ gauge boson, is given by 
\begin{eqnarray} 
\gamma &=&'3 \bar\nu_l \gamma_\mu \nu_l - \bar q_1 \gamma_\mu q_1 
-\bar q_2 \gamma_\mu q_2-\bar q_3 \gamma_\mu q_3'  
\label{eqg}
\end{eqnarray}
A straightforward calculation yields 
\begin{eqnarray} 
\gamma &=& f^*_0 (f_1 + f_2 +f_3)
+f^*_1 f_2 +f^*_1 f_3 +f^*_2 f_3  +c.c.
\label{egamk}
\end{eqnarray}
where $f_0=\phi_{\overline{1234}}$, $f_1=\phi_{\overline{2341}}$, 
$f_2=\phi_{\overline{3412}}$ and $f_3=\phi_{\overline{4123}}$ for 
$Z_4$. In the case of $K$ the result eq. (\ref{egamk}) is 
formally identical with $f_1$ and $f_3$ replaced by 
$f_1=\phi_{\overline{2143}}$ and $f_3=\phi_{\overline{4321}}$. 

One may compare this result with the expression for the 
$U(1)_{B-L}$-photon obtained in the weak sector  
\begin{eqnarray} 
\gamma&=&' \bar\nu_l \gamma_\mu \nu_l + \bar l \gamma_\mu l ' \nonumber \\
&=& 2[ \phi^*_{id} \phi_{id} + \phi^*_{u_0} \phi_{u_0}] 
\label{eq2xx2}
\end{eqnarray}
to see that the full $S_4$-expression for the photon will 
contain the terms eq. (\ref{egamk}) together with corresponding 
odd $(1\leftrightarrow 3)$ contributions. 

Next the diagonal gluons are given by 
\begin{eqnarray} 
g_3 &=&' \bar q_1 \gamma_\mu q_1 -\bar q_3 \gamma_\mu q_3 '  \nonumber \\
&=& \frac{1}{2} [\mp f^*_0 (f_1 -f_3) -f^*_1 f_2 \mp f^*_2 f_3] \mp c.c. 
\label{eqg3} \\
g_8 &=&'2\bar q_2 \gamma_\mu q_2 
-\bar q_1 \gamma_\mu q_1-\bar q_3 \gamma_\mu q_3'      \nonumber \\
&=&\frac{1}{2} [f^*_0 (2f_2-f_1-f_3)+ 2f^*_1 f_3-f^*_1 f_2 -f^*_2 f_3]+c.c.
\label{eqg8}
\end{eqnarray}
where the upper sign in eq. (\ref{eqg3}) corresponds to 
$Z_4$ and the lower to $K$. 
The results eqs. (\ref{egamk}), (\ref{eqg3}) and (\ref{eqg8}) 
can be directly compared to the charge matrices 
eqs. (\ref{eq7145595})-(\ref{eq9209}) obtained at the 
end of the last section. 

Analogously one gets the nondiagonal gluons like 
\begin{eqnarray} 
g_{13} &=& '\frac{1}{2} [ \bar q_1 \gamma_\mu q_3 +\bar q_3 \gamma_\mu q_1] ' \nonumber  \\
&=& \frac{1}{4} \{|f_0|^2 - |f_1|^2 +|f_2|^2 - |f_3|^2 
                  +[f^*_1 f_3 -f^*_0 f_2 +c.c.]\} \\
g_{31} &=&' \frac{1}{2} [\bar q_1 \gamma_\mu q_3 -\bar q_3 \gamma_\mu q_1] '  \nonumber \\
&=& \frac{1}{2} [\pm f^*_1 f_2 -f^*_2 f_3 -f^*_0 f_1 +f^*_0 f_3 \pm c.c.] 
\label{eq24g8}
\end{eqnarray}
where again the upper sign is for $Z_4$ and the lower for $K$. 

Unfortunately on this level leptoquarks $\bar l \gamma_\mu q_i$ 
exist as well. For example, one has 
\begin{eqnarray} 
LQ_{02} &=& '\frac{1}{2} [ \bar l \gamma_\mu q_2 +\bar q_2 \gamma_\mu l] '  \nonumber \\
&=& \frac{1}{4} \{|f_0|^2 - |f_1|^2 +|f_2|^2 - |f_3|^2 
                  -[f^*_1 f_3 -f^*_0 f_2 +c.c.]\} \\
LQ_{20} &=&' \frac{1}{2} [\bar l \gamma_\mu q_2 -\bar q_2 \gamma_\mu l] '  \nonumber \\
&=& \frac{1}{2} [f^*_1 f_2 -f^*_2 f_3 -f^*_0 f_1 -f^*_0 f_3 - c.c.] 
\label{eq111g8}
\end{eqnarray}
In order that they vanish in accord with 
the nonexistence theorem and with phenomenology one has to impose 
the following additional constraints    
\begin{eqnarray} 
|f_1+f_2|^2 = |f_0+f_3|^2 & &\\
|f_1+f_3|^2 = |f_0+f_2|^2 & &\\
|f_2+f_3|^2 = |f_0+f_1|^2 & &\\
f^*_0 f_1 -f^*_1 f_2 -f^*_1 f_3 - c.c. &=& 0 \\
f^*_0 f_2 +f^*_1 f_2 -f^*_2 f_3 - c.c. &=& 0 \\
f^*_0 f_3 +f^*_1 f_3 +f^*_2 f_3 - c.c. &=& 0 
\label{eqsca}
\end{eqnarray}
A straightforward calculation then yields 
$\bar l \gamma_\mu q_1 =\bar l \gamma_\mu q_2 =\bar l \gamma_\mu q_3 =0$ 
which is the desired result. Note that the absence of 
leptoquarks indicates that the effective 
gauge group is not $SU(4)$ but $U(1)_{B-L}\times SU(3)_c$. 
There is then no room in this model for exotic grand unified 
vector bosons. 

Imposing the constraints on the other vector 
bosons allows to represent them in a somewhat 
more convenient form. For example, for the photon one obtains 
\begin{eqnarray} 
\gamma &=& 3|f_0|^2-|f_1|^2-|f_2|^2-|f_3|^2+ [2f_0^* (f_1+f_2+f_3)+c.c.]
\label{eq2xx2}
\end{eqnarray}


One may extend this analysis step by step to $D_4$- and then 
to $S_4$-functions. Applying symmetry under 
$\overline{3214}=(1\leftrightarrow 3)$ (plus $Z_3$-family transformtations) 
yields gluons and photon whereas symmetry under even 
permutations $g \in A_4$ yields the weak gauge bosons. 


In summary we have analyzed the posible interactions 
between tetrahedron and antitetrahedron states and 
shown which transformations should be symmetries of the 
resulting vector boson state and which ones should not. 
As a consequence, the 24$^2$ $S'_4 \times S_4$ 
tetrahedron-antitetrahedron states get reduced to 
\begin{itemize}
\item
4 states $\gamma$, $W^\pm$ and $Z$ 
generated by 
\begin{eqnarray} 
S'_4/A'_4 \times S_4/(Z_4\diamond Z_3) = Z'_2 \times Z_2
\label{eqgg4}
\end{eqnarray}
where $A_4 \subset S_4$ is the group of even 
permutations in $S_4$. The first factor in (\ref{eqgg4}) 
is due to the symmetry under even 
transformation / symmetry breaking under odd transformations 
whereas the 
second factor corresponds to the 12 orbits in the family 
and color group. In more concrete terms the resulting 4 
real degrees of freedom are given by 
\begin{eqnarray} 
|\phi_{\overline{1234}}|^2 & & \nonumber \\
|\phi_{\overline{3214}}|^2 & & \nonumber \\
\phi_{\overline{1234}}^* \phi_{\overline{3214}} 
&=&[\phi_{\overline{3214}}^* \phi_{\overline{1234}}]^*  
\end{eqnarray}
objects, which have already appeared, in a somewhat different 
notation, in eqs. (\ref{eqw5}) and (\ref{eqzxx2}). Note 
that the 'singlet' 
\begin{eqnarray} 
|\phi_{id}|^2 + |\phi_{u_0}|^2=
|\phi_{\overline{1234}}|^2 + |\phi_{\overline{3214}}|^2 
\end{eqnarray} 
corresponds to the photon. As we have seen, for the photon 
this is not the whole story, because the 
factorization of the photon interaction takes 
place both in the weak and the strong sector. 
\item
16 gluon and leptoquark states generated either by $K' \times K$ 
or $Z'_4 \times Z_4$ 
and in more concrete terms given by $f_i^* f_j$, 
precisely the objects which appear in the above equations for 
photons and gluons.
\end{itemize}

There is one item which has been spared out so far: the question of 
spin and the related topic of parity violation. 
In fact, most of the preceeding analysis can be carried out 
for lefthanded and righthanded fermions separately. This way 
one naturally ends up with separate weak bosons for left 
and right. 
For the weak interactions this is okay, because we are expecting 
an effective $SU(2)_L \times SU(2)_R$ gauge theory. Photon 
and gluons, however, behave differently. They are identical for 
left- and righthanded fermions, i.e. they interact vectorlike. 
The question then immediately arises: what makes them peculiar?

To answer this question it is not necessary to know all about 
the spin properties of the tetrons themselves. What one 
has to realize is, that wave functions $\phi_u$ for odd 
permutations $u \in S_4$ naturally have opposite helicity 
to wave functions $\phi_g$ for odd permutations $g \in S_4$ 
and therefore should enter the linear combinations 
for fermions of definite helicity with a parity operator 
in front (assuming relativistic motion). 
The point is that, considered as an $O(3)$-transformation, an odd 
permutation has $det(u)=-1$ and therefore intrinsically 
contains a parity transformation which has to be 
taken care of. For definiteness one may consider the 
$K$ wave functions eqs. (\ref{eq46700})-(\ref{eq4670}). 
Eq. (\ref{eq46700}) has to be modified according to 
\begin{equation} 
\phi_{\nu_\tau} =\frac{1}{\sqrt{4}}[\phi_{\overline{1234}} 
+ P \phi_{\overline{2341}} + \phi_{\overline{3412}} + 
P\phi_{\overline{4123}}] 
\label{eq455555}
\end{equation}

In the simple case eq. (\ref{eq4833}) of just 2 isospin partners 
$\phi_{\pm}$ the four helicity states would be given by 
\begin{eqnarray} 
\phi_{+,L}&=&\frac{1}{\sqrt{2}}[\phi_{\overline{13}}+P\phi_{\overline{31}}] \\
\phi_{-,L}&=&\frac{1}{\sqrt{2}}[\phi_{\overline{13}}-P\phi_{\overline{31}}]  \\
\phi_{+,R}&=&\frac{1}{\sqrt{2}}[P\phi_{\overline{13}}+\phi_{\overline{31}}] \\
\phi_{-,R}&=&\frac{1}{\sqrt{2}}[P\phi_{\overline{13}}-\phi_{\overline{31}}]
\label{eq45x1x}
\end{eqnarray} 

Similarly, in the general case eq. (\ref{eq500})ff of $S_4$-functions 
left handed fermion states will be generically of the form 
\begin{equation} 
\phi_L=g_\phi + Pu_\phi
\label{eq45152}
\end{equation}
i.e. linear combinations of even and $P\circ$odd contributions, 
whereas right handed fermions will have the form 
\begin{equation} 
\phi_R=Pg_\phi + u_\phi
\label{eq45153}
\end{equation} 
Applying an even transformation $g\in S_4$ to the 
spin-1 object 
\begin{eqnarray} 
\bar \phi_L \gamma_\mu \phi_L &=& \bar g_\phi \gamma_\mu g_\phi 
+ \bar{Pu_\phi} \gamma_\mu Pu_\phi + \bar g_\phi \gamma_\mu Pu_\phi 
+ \bar{Pu_\phi} \gamma_\mu g_\phi \nonumber \\
&\rightarrow_g & \bar g'_\phi \gamma_\mu g'_\phi 
+ \bar{Pu'_\phi} \gamma_\mu Pu'_\phi + 
\bar g'_\phi \gamma_\mu Pu'_\phi + 
\bar{Pu'_\phi} \gamma_\mu g'_\phi
\label{eq45154}
\end{eqnarray} 
where this time I have included the bar and the $\gamma$-matrix,
does not change the helicity, i.e. a weak vector boson $W_L$ which 
according to eq. (\ref{eqgg4}) is obtained by even permutations 
as symmetry transformations remains lefthanded. There is no 
relation between $\bar \phi_L \gamma_\mu \phi_L$ and 
$\bar \phi_R \gamma_\mu \phi_R$, i.e. $W_L$ and $W_R$ 
are completely independent objects, i.e. both of the standard 
form eq. (\ref{eqzxx2}) with index L and R, respectively. 

In contrast, applying an odd transformation $u\in S_4$ 
like $\overline{3214}=(1\leftrightarrow 3)$, which is the 
symmetry under which the gluons and the photon remain 
invariant, one ends up with a right handed vector boson 
\begin{eqnarray} 
\bar \phi_L \gamma_\mu \phi_L &=& \bar g_\phi \gamma_\mu g_\phi 
+ \bar{Pu_\phi} \gamma_\mu Pu_\phi + \bar g_\phi \gamma_\mu Pu_\phi 
+ \bar{Pu_\phi} \gamma_\mu g_\phi \nonumber \\
&\rightarrow_u & \bar u''_\phi \gamma_\mu u''_\phi 
+ \bar{Pg''_\phi} \gamma_\mu Pg''_\phi 
+ \bar u''_\phi \gamma_\mu Pg''_\phi + 
\bar{Pg''_\phi} \gamma_\mu u''_\phi  \nonumber \\
&=& \bar \phi_R \gamma_\mu \phi_R
\label{eq45154}
\end{eqnarray} 
This means imposing this symmetry instantly leads to a vectorlike 
structure of photon and gluons. 



\section{Conclusions}

In conclusion we have here a scheme which accomodates all 
observed fermions and vector bosons. In addition it relates 
the number of these states in an obvious and natural way to the 
number of space dimensions, because the tetrahedron with 4 fundamental 
constituents is the minimal complex to build up 3 dimensions. 
With 3 constituents, for example, one would live in 
a 2-dimensional world with bound states of triangles 
which would yield only 6 bound permutation states instead of 24. 

As we have seen, there are several other respects like parity 
violation and mass hierarchies in which the present model 
goes a step further in understanding than standard gauge theories. 
The real challenge will of course be to understand the nature 
of the tetron interactions and to write them in a 
renormalizable form. 

About possible experimental tests: in the low energy limit the 
tetrahedron goes over into a point, i.e. it becomes an ordinary pointlike 
fermion. Increasing the energy one should be able to dissolve 
its spatial extension which will show up e.g. in the form of 
non-Dirac-like form factors. The question then is: how 
small are the extensions of the tetrahedron? This question is 
difficult to answer and is related to the problem of the 
strength and nature of the superstrong interaction which 
binds the tetrons together. Certainly there will be a correspondence 
between the tetrahedron extension and the new coupling 
constant which is a new fundamental constant of nature from which 
most of the couplings and masses known in particle physics will be 
derived. 

\section{Appendix: Group and Representation Structure of $T_d=S_4$}

$S_4$ is the group of permutations of 4 objects and isomorphic 
to the symmetry group $T_d$ of an equilateral tetrahedron. 
Among the subgroups of $S_4$ there are $A_4$, $Z_3$, $S_2$, $S_3$, $Z_4$ 
and the so-called $Kleinsche$ $Vierergruppe$ K consisting of permutations 
$\overline{1234}$, $\overline{2143}$, $\overline{3412}$ and $\overline{4321}$ 
which is isomorphic to $Z_2 \times Z_2$. It is 
is the smallest noncyclic group and describes the 
symmetries of a rectangle in 2 dimensions. 
$Z_3$ is the cyclic group of 3 elements or equivalently the 
subgroup of $S_3$ corresponding to the even permutations 
or equivalently the group generated by $\pm 2\pi/3$ rotations 
of the plane. 

$S_4$ can be written as $Z_4 \diamond S_3$ where $S_3=Z_3 \diamond S_2$ is the point 
symmetry group of an equilateral triangle. Correspondingly there are 
4 orbits of $S_3$ and 6 orbits of $Z_4$ in $S_4$. 
In the present model the 4 degrees of $Z_4$ correspond to a lepton 
and a quark with three colors. $Z_3$ is the family group and 
$S_2$ roughly corresponds to weak isospin. 

A short review of the representations of $S_4$ is in order: 
for a finite group the number of irreducible representations is 
given by the number of conjugacy classes (=5 in this case). Apart from 
the trivial representation $A_1$ there is the totally antisymmetric 
representation $A_2$, which assigns 1 to all even and -1 to every 
odd permutation. $A_1$ and $A_2$ are of dimension 1. There are 
two 3-dimensional representations usually called $T_1$ and $T_2$, 
$T_1$ describing the action of the permutation group on the tetrahedron. 
$T_2$ differs from $T_1$ 
by a minus sign for odd permutations. Finally there is a 2-dimensional 
representation $E$ which may be considered as the trivial extension 
of a corresponding 2-dimensional representation of $S_3$. 

One may use the representations $E$ and $T_1$ in the 
context of the symmetry breaking, because in first order 
approximation the representation matrices may be considered as state 
vectors (in the case of $E$ an average 'family state vector') of 
the fermions. This has to do with the fact that the permutations 
(and its representations) may be considered as ladder operators 
which generate all states $\overline{abcd}$ from the ground state 
$\overline{1234}$. These unbroken states can then be used, in the 
spirit of first order perturbation theory, to derive mass and 
charge relations for the fermions. (I have not used this approach 
in the present paper, because I found a more elegant way to 
derive these relations.) 

The covering group $\tilde S_4$ of $S_4$ is embedded in $SU(2)$ just 
as $S_4$ is in $SO(3)$. Apart from those 
representations which are extensions from 
$S_4$ to $\tilde S_4$ it has the following representations: 
a 2-dimensional representation $G_1$ which is the restriction of 
the fundamental representation of $SU_2$ to the symmetry 
transformations of a tetrahedron in much the same way as $T_1$ 
is the restriction of the fundamental representation of $SO_3$. 
Then there is a representations $G_2$ 
obtained from $G_1$ like $T_2$ is obtained from $T_1$ 
and a 4-dimensional (spin 3/2) representation H.


\end{document}